\documentclass[twocolumn,preprintnumbers,amsmath,amssymb,nofootinbib]{revtex4}

\usepackage{graphicx}
\usepackage{bm}

\newcommand{\ee}{\begin{equation}}
\newcommand{\eee}{\end{equation}}
\newcommand{\ea}{\begin{eqnarray}}
\newcommand{\eea}{\end{eqnarray}}

\newcommand{\mplank}{\textrm{M}_{\textrm{P}}}

\newcommand{\CMBFAST}{{\sc cmbfast}}
\newcommand{\CAMB}{{\sc camb}}
\newcommand{\CMBEASY}{{\sc cmbeasy}}

\begin{document}

\author{Michael Doran}
\affiliation{Institut f\"ur  Theoretische Physik, Philosophenweg 16, 69120 Heidelberg, Germany}
\affiliation{Department of Physics \& Astronomy,
        HB 6127 Wilder Laboratory,
        Dartmouth College,
        Hanover, NH 03755, USA}

\preprint{HD-THEP-05-03}
\title{Speeding Up Cosmological Boltzmann Codes}

\begin{abstract}
We introduce a novel  strategy for cosmological Boltzmann codes leading to
an increase in speed by a factor of $\sim 30$ for small scale Fourier modes. We
(re-)investigate the tight coupling approximation and obtain analytic formulae including
the octupoles of photon intensity and polarization. 
Numerically, these results reach optimal precision.
Damping rapid oscillations of small scale modes at later times, 
we simplify the integration of cosmological perturbations. We obtain
analytic expressions for the photon density contrast and velocity as well as an estimate
of the quadrupole from after last scattering until today. These analytic formulae
hold well during re-ionization and are in fact negligible for realistic 
cosmological scenarios. However, they do extend the validity of our approach
to models with very large optical depth to the last scattering surface. 
\end{abstract}
\maketitle

\newcommand{\mpc}{\,\textrm{Mpc}}
\newcommand{\cpm}{\mpc^{-1}}

\newcommand{\kb}{10^{-2} \cpm}
\newcommand{\delb}{\delta_b}
\newcommand{\delg}{\delta_\gamma}
\newcommand{\ddelb}{\dot \delta_b}
\newcommand{\ddelg}{\dot \delta_\gamma}

\newcommand{\tc}{\tau_c}
\newcommand{\vb}{v_b}
\newcommand{\vg}{v_\gamma}
\newcommand{\dvb}{\dot v_b}
\newcommand{\dvg}{\dot v_\gamma}
\newcommand{\ddvb}{\ddot v_b}
\newcommand{\ddvg}{\ddot v_\gamma}
\newcommand{\adota}{\frac{\dot a}{a}}
\newcommand{\addota}{\frac{\ddot a}{a}}

\newcommand{\csg}{c_s^\gamma}
\newcommand{\cs}{c_s^2}
\newcommand{\dcs}{\dot c_s^2}
\newcommand{\dtc}{\dot \tau_c}
\newcommand{\tci}{\tau_c^{-1}}
\newcommand{\M}{\mathcal{M}}
\newcommand{\N}{\mathcal{N}}
\newcommand{\C}{\mathcal{C}}
\newcommand{\sg}{\sigma_\gamma}
\newcommand{\dsg}{\dot \sigma_\gamma}
\newcommand{\kt}{k\tc}
\newcommand{\bkt}{\left(k\tc\right)}
\newcommand{\slip}{\mathcal{\dot V}}
\section{Introduction}
Standard codes such as \CMBFAST\ \cite{Seljak:1996is,cmbfast}, \CAMB\ 
\cite{Lewis:1999bs,Lewis:2002ah,camb} 
or \CMBEASY\ \cite{Doran:2003sy,Doran:2003ua,cmbeasy} compute the evolution of 
small perturbations in a Friedman-Robertson-Walker Universe. The output most
frequently used are multipole spectra of the Cosmic Microwave Background (CMB)
and power spectra of massive particles. These predictions are compared
to  precision measurements of the Cosmic 
Microwave Background (CMB) \cite{Spergel:2003cb} and
Large Scale Structure (LSS) \cite{Tegmark:2003ud}. 
Working in Fourier space,  the codes evolve perturbation equations for single Fourier $k$-modes.
The simulated evolution starts well outside the horizon at early times and ends today.
For the CMB, relevant scales lie in the range $k \sim 10^{-5} \dots 1 \cpm$,
while those for the LSS extend to higher $k \sim 5 \dots 1000 \cpm$.

Currently, the time needed to evolve a single mode is roughly proportional to $k$. 
As the spectrum is computed in logarithmic $k$-steps, the largest
few $k$-modes tend to dominate the resources needed for the entire calculation.
We have analyzed the current strategy to integrate the perturbation equations and 
singled out two bottlenecks. The first one is the so called tight 
coupling regime (or better: the end of  tight coupling).
The second one are rapid oscillations of relativistic quantities for high $k$-modes.
Roughly speaking, in standard \CMBFAST\ and \CMBEASY, both regimes contribute 
equally to the computational cost. This is likely not the case for \CAMB, as it
uses a higher order scheme during tight coupling -- a solution similar\footnote{To our
knowledge, there is no published discussion on higher order schemes. There is,
however unpublished work by Antony Lewis and Constantinos Skordis \cite{constantinos}.}
 to the one we will present later on.

\begin{figure}
\includegraphics[width=0.42\textwidth]{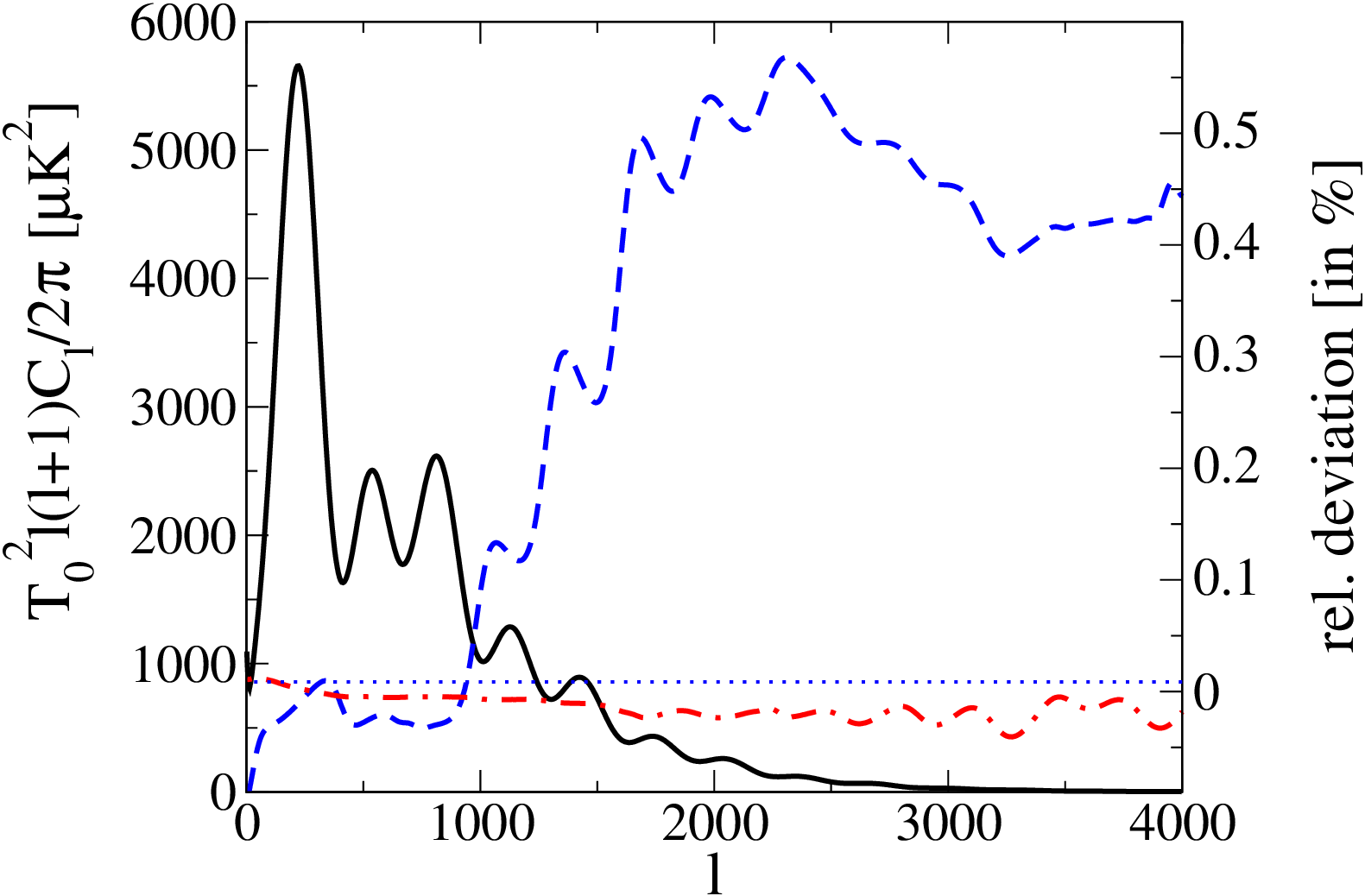} 
\caption{The CMB multipole spectrum up to $l=4000$ for a standard cosmological 
model (solid line). 
The dashed (blue) line shows the relative deviation between a standard \CMBEASY\ calculation
and one where the switch ending tight coupling has been pushed to earlier times and hence better
precision. The geometric average deviation is $\sim 0.3\%$. The dashed-dotted (red) line shows
the deviation between such a high precision \CMBEASY\ calculation and the new algorithm. With the
average geometric deviation $\sim 0.01\%$ roughly 30 times smaller, our new algorithm comes
close to the optimal result.}\label{fig::differences}
\end{figure}

Our strategy therefore consists of two parts. The first one
is a revised tight coupling treatment. In this, we will make a \emph{conceptual} change, 
distinguishing
between tight coupling of the baryon and photon fluid velocities on one hand and the validity 
of an analytic treatment of the photon intensity and 
polarization quadrupole  on the other. 
In essence, our solution extends to the octupole. We thus
capture the physics during tight coupling better than previously achieved. 
This leads to a considerable increase in accuracy reaching the optimal precision for this 
stage of the computation (see Figure \ref{fig::differences}).

The second part of our solution consists of suppressing unwanted oscillations in
the multipole components of relativistic particles. 
In essence, it is the line-of-sight \cite{Seljak:1996is}
formulation of all modern CMB codes that allows us to do this. As we will see, the 
oscillations we suppress are anyhow unphysical as they 
perpetuate unwanted reflections due to truncation effects.
In any case, the modifications are such that observational quantities like the CMB or LSS
are not influenced by our choice.
These two improvements combined lead to considerably shorter integration times. 
Typically, the benefit sets in for modes $k \gtrapprox 0.1 \cpm$ and increases
gradually until reaching factors of $\sim 30$ for modes $\sim 5\cpm$ and higher. 
For some speed comparisons, see Table \ref{tab::speed}.

\begin{table}
\begin{ruledtabular}
\begin{tabular}{lccc}
$k_{max}/h$ & $l_{max}$ & \CMBEASY& \CMBEASY\\%
&  & (new algorithm) & (sync. gauge) \\%
$10\cpm$ & no CMB & 1.5s & 10s\\ %
$100\cpm$ & no CMB& 4s & 93s \\
$5\cpm$ &2000 & 5s & 12s\\ 
$10\cpm$ & 4000  & 9s & 25s %
\end{tabular}
\end{ruledtabular}
\caption{Comparison of speed between the new algorithm and the standard synchronous 
gauge implementation. Execution times of \CMBFAST\ are comparable to the standard
synchronous gauge implementation, but can deviate by a factor of $\sim \frac{1}{2} \dots 2$ 
from \CMBEASY\ depending on the task. The Hubble parameter for the model used was $h=0.7$.}\label{tab::speed}
\end{table}

\section{Tight coupling revised}
\begin{figure}
\includegraphics[width=0.38\textwidth]{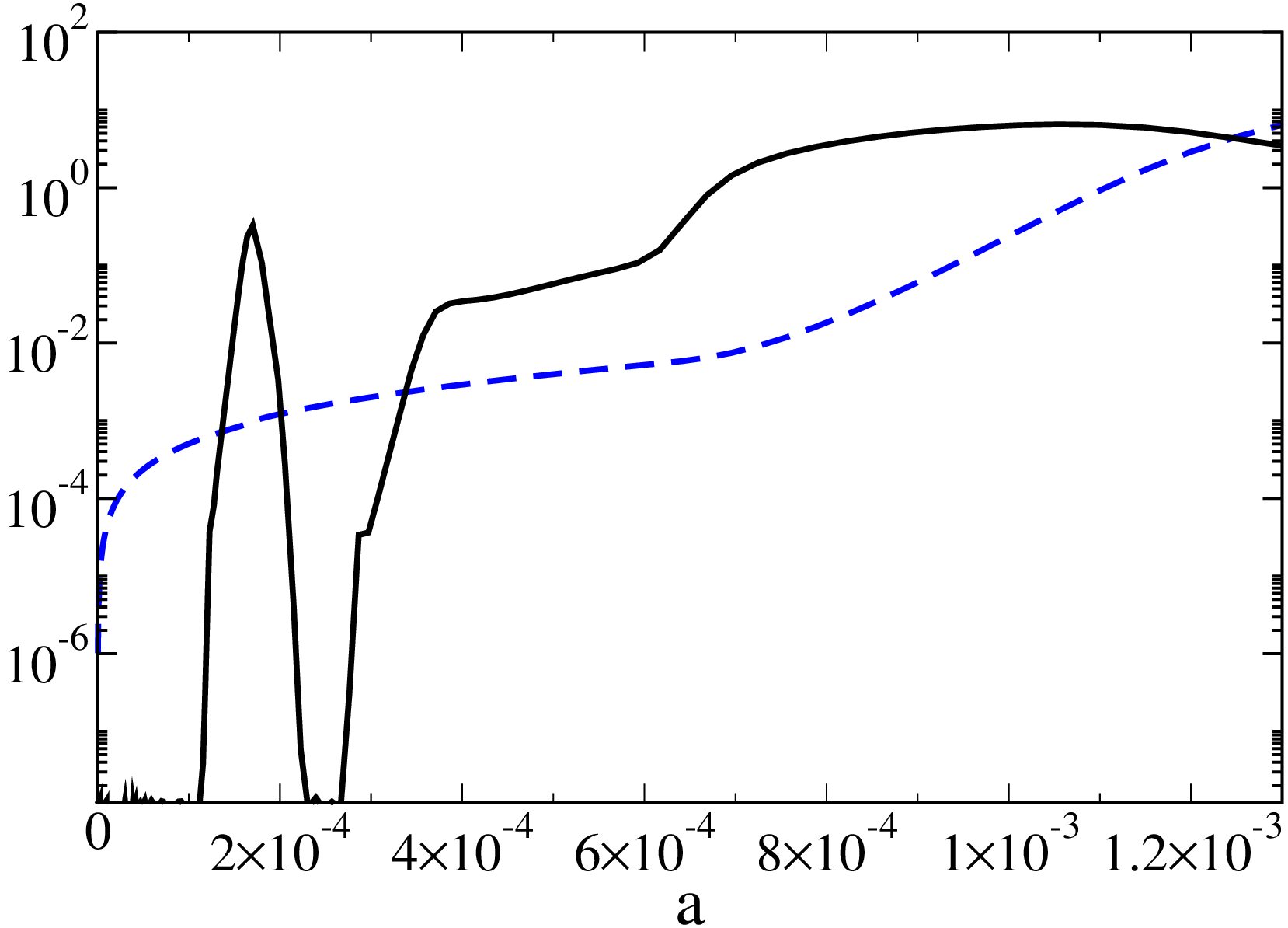}
\caption{Relative deviation of $\dot \tc$  from the naive scaling relation:
$[\dot \tc - 2  (\dot a /a) \tc]  / [ 2  (\dot a /a) \tc]$ (solid line). We also depict 
the product $\tc \adota$ (dashed line) vs. the scale factor $a$, which compares the 
mean free path to the expansion rate of the Universe.
In the cosmological model used, matter radiation equality is at $a_{equ} = 3\times 10^{-4}$ 
and last scattering defined by the peak of the visibility function is at $a_{ls} = 9 \times 10^{-4}$. 
The deviation around $a = 2 \times 10^{-4}$ is from Helium recombination and is practically negligible, 
because the visibility is still small during that period. At later times, however the deviation is due to the onset
of Hydrogen recombination and takes on substantial values before last scattering. } \label{fig::scaling}
\end{figure}

At early times, the photon and baryon fluids are strongly coupled via Thomson
scattering. The mean free path between collisions of a photon $\tci \equiv a n_e \sigma_{T}$ is
given in terms of the number density of free electrons $n_e$, the scale factor of 
the Universe $a$ and Thomson cross section $\sigma_{T}$. During early times, 
 Hydrogen and Helium are fully ionized, hence $n_e \propto a^{-3}$ and $\tc \propto a^{2}$.
During Helium and Hydrogen recombination, this scaling argument does not hold 
(see Figure \ref{fig::scaling}). To avoid these  periods 
we resort to 
the correct value of $\dtc$ computed beforehand instead of using $\dtc = 2\adota \tc$ 
for redshifts $z < 10^{4}$. The effect of  assuming that the scaling holds would however 
be considerably less than $1\%$ on the  final CMB spectrum.

To discuss the tight coupling regime, let us 
recapitulate the evolution equations for baryons and photons. We do this  
in terms of their density
perturbation $\delta$ and  bulk velocity $v$. For photons, we additionally consider the
shear $\sigma_\gamma$ and higher multipole moments $\M_l$ of the intensity as well 
as polarization multipoles $E_l$. Our variables are related to the ones of \cite{Ma:1995ey} by
substituting $v \to k^{-1} \theta$. 
In longitudinal gauge, baryons evolve according to
\ea
\ddelb &=& -k \vb + 3 \dot \phi\\
\dvb &=& -\adota \vb + \cs k \delb + R \tci (\vg - \vb) + k\psi \label{eqn::dvb},
\eea
where $R \equiv (4/3) \rho_\gamma / \rho_b$, the speed of sound of the baryons
is denoted by $\cs$ and $\phi$ and $\psi$ are metric perturbations. 
By definition, $R \propto a^{-1}$ (provided no baryons are converted to other forms of
energy) and at the time of interest, $\cs \propto T_b = T_\gamma \propto a^{-1}$
(for more detail see e.g. \cite{Ma:1995ey}).
Photons evolve according to the hierarchy 
\ea
\ddelg &=& -\frac{4}{3} k \vg + 4 \dot \phi \label{eqn::ddelg}\\
\dvg &=& k\left(\frac{1}{4}\delg - \sg\right) + k\psi + \tci (\vb - \vg), \label{eqn::dvg}\\
\frac{5}{2} \dsg &=& \dot \M_2 = -\tci \left( \frac{9}{10} \M_2 + \frac{\sqrt{6}}{10} E_2 \right) \nonumber \\ &&\hspace{2.5cm}
+ k\left(\frac{2}{3}\vg - \frac{3}{7}\M_3 \right)  \label{eqn::m2} \\
\dot \M_l &=&  k\! \left( \frac{l}{2l -1} \M_{l-1}\! -\! \frac{l+1}{2l+3} \M_{l+1}\right)\!-\!\tci \M_l, \label{eqn::ml}
\eea
where the $E$-type polarization obeys 
\ea
\dot E_2 &=& -k \frac{\sqrt{5}}{7} E_3 - \tci \left(\frac{4}{10} E_2 + \frac{\sqrt{6}}{10}M_2 \right) \label{eqn::e2} \\
\dot E_l &=& - \tci E_l \nonumber \\ 
&& \ \
+k\left(\frac{\sqrt{l^2-4}}{2l-1} E_{l-1} - \frac{\sqrt{l^2+2l-3}}{2l+3}E_{l+1}\right).  \label{eqn::el}
\eea
The overwhelmingly large value of $\tci$ precludes a straight forward
numerical integration at early times: tiny errors in the propagation
of $\vb$ and $\vg$ lead to strong restoring forces. This severely
limits the maximum step size of the integrator and hence the speed of
integration. Ever since Peebles and Yu \cite{Peebles:ag} first
calculated the CMB fluctuations, one resorts to the so called tight coupling
approximation. This approximation eliminates all terms of order $\tci$
from the evolution equations assuming\footnote{There is no restoring force left, as we will
see. Any error in the approximation is therefore amplified over time. One could, in principle
retain a fraction of the restoring force to eliminate small numerical errors. However,
this is not necessary in practice and we therefore will not discuss this possibility further.}
 tight coupling at initial
times. Our discussion will closely lean on that of 
\cite{Ma:1995ey}, taking a slightly different route. In contrast
to \cite{Ma:1995ey}, however, we will  keep \emph{all} terms in the
derivation.
Like \cite{Ma:1995ey}, we start by  solving \eqref{eqn::dvg} for $(\vb - \vg)$ and 
write $\dvg = \dvb + (\dvg - \dvb)$ to get Equation (71) of \cite{Ma:1995ey}
\ee\label{eqn::slip71}
(\vb - \vg) = \tc\left[\dvb + (\dvg - \dvb) - k\left(\frac{1}{4}\delg - \sg + \psi  \right)\right].
\eee 
Substituting Equation \eqref{eqn::dvb} for $\dvb$ into this Equation
\eqref{eqn::slip71}, one gets Equation (72) of \cite{Ma:1995ey}
\begin{multline}\label{eqn::slip72}
\frac{(1+R)}{\tc}(\vb - \vg) = -\adota \vb + (\dvg - \dvb) \\
+ k\left(\cs \delb - \frac{1}{4}\delg + \sg \right).
\end{multline}
Deriving the LHS of this Equation \eqref{eqn::slip72} yields
\ea
\dot{\textrm{lhs}}  &=& \frac{(1+R)}{\tc}(\dvb - \dvg)  \nonumber\\
 && \quad - (\vb-\vg)    \left[ \adota \frac{R}{\tc} - \frac{1+R}{\tc}\frac{\dtc}{\tc} \right]  \\
&\overset{*}{=}&   \frac{(1+R)}{\tc}(\dvb - \dvg) -  \frac{2+3R}{\tc}\adota(\vb - \vg),
\eea
where the last line holds provided the assumed scaling of $\tc$ is correct (see also
Figure \ref{fig::scaling}).
All in all, deriving  Equation \eqref{eqn::slip72} with respect to conformal time yields
\begin{multline}\label{eqn::slip0}
\frac{(1+R)}{\tc}(\dvb - \dvg) - \left[ \adota \frac{R}{\tc} - \frac{1+R}{\tc}\frac{\dtc}{\tc} \right]  (\vb-\vg)      \\ =  (\ddvg - \ddvb)
-\addota \vb +\left(\adota\right)^2\vb 
- \adota \dvb \\ + k\left(\dcs \delb + \cs\ddelb - \frac{1}{4}\ddelg +  \dsg  \right )
\end{multline}
Multiplying Equation \eqref{eqn::dvb} by $\adota$ to substitute $\adota \dvb$ in \eqref{eqn::slip0}, we get
\begin{multline}\label{eqn::slip1}
\frac{(1+R)}{\tc}(\dvb - \dvg)  = \left[ \adota \frac{R}{\tc} - \frac{1+R}{\tc}\frac{\dtc}{\tc} \right]  (\vb-\vg)   \\ 
+ (\ddvg - \ddvb) -\addota \vb +2 \left(\adota\right)^2\vb - 2 \adota  \cs k \delb  \\ 
 + k\left(\cs\ddelb - \frac{1}{4}\ddelg +  \dsg \right ) + \adota k \psi
\end{multline}
where we have used $\dcs = -\adota \cs$. We could stop here, however it is numerically better conditioned
to write $2\left(\adota\right)^2\vb =2 \adota \left(\adota \vb\right)$ where $\adota \vb$ is obtained
from solving Equation \eqref{eqn::slip72} for $\adota \vb$. This expression for $\left(\adota\right)^2\vb$ is then plugged into Equation \eqref{eqn::slip1} to yield 
the final result for the slip (denoted by $\mathcal{\dot V}$)
\begin{multline}\label{eqn::slip74}
\mathcal{\dot V} \equiv (\dvb - \dvg) = \Bigg \{ \left[\frac{\dtc}{\tc} - \frac{2}{1+R}\right]\adota(\vb - \vg) \\+ \frac{\tc}{1+R} \bigg[  -\addota \vb   +   (\ddvg - \ddvb) + k \left( \frac{1}{2} \delg - 2\sg + \psi\right) \\
+   k\left(\cs\ddelb - \frac{1}{4}\ddelg +  \dsg  \right )
\bigg] \Bigg\} \Bigg / \left\{1+2 \adota \frac{\tc}{1+R}\right\}.\end{multline}
or alternatively, at times when the scaling of $\tc$ holds, 
\begin{multline}\label{eqn::slip74b}
\mathcal{\dot V} \equiv (\dvb - \dvg) = \Bigg \{  \frac{2R}{1+R}\adota(\vb - \vg) \\+ \frac{\tc}{1+R} \bigg[  -\addota \vb   + (\ddvg - \ddvb) + k \left( \frac{1}{2} \delg - 2\sg + \psi\right) \\
+   k\left(\cs\ddelb - \frac{1}{4}\ddelg +  \dsg  \right )
\bigg] \Bigg\} \Bigg/  \left\{1+2 \adota \frac{\tc}{1+R}\right\}.\end{multline}
This Equation \eqref{eqn::slip74} (or more obviously \eqref{eqn::slip74b}) is essentially Equation (74) of \cite{Ma:1995ey} up to some corrections. Having kept all terms, we note
that our Equation \eqref{eqn::slip74} is \emph{exact}. To obtain Equations of motion for $\vb$ and $\vg$ 
during tight coupling, we plug our result for $(\dvb - \dvg)$, Equation \eqref{eqn::slip74} into the RHS of 
Equation \eqref{eqn::slip71} and this in turn into the RHS of Equations \eqref{eqn::dvb} and \eqref{eqn::dvg}.
This yields
\ea
\dvb &=& \frac{1}{1+R}\left(  k \cs \delb - \adota\vb    \right) +  k\psi \nonumber  \\
&& \qquad\frac{R}{1+R}\left[k\left(\frac{1}{4}\delg - \sg \right) + \slip\right] \nonumber \\
\dvg &=& \frac{R}{1+R}k \left(\frac{1}{4}\delg - \sg \right) +k \psi \nonumber \\ 
&& \qquad + \frac{1}{1+R}\left(k \cs\delb - \adota \vb -\slip       \right) 
 \label{eqn::tightdv}
\eea
Up to now, we have made no approximations. Conceptually, we would like to 
separate the question of tight coupling for the 
velocities $\vg$ and $\vb$ from any approximations of 
the shear $\sg$ which we make below. 
As far as the tight coupling of the velocities and hence the slip $\slip$ is concerned,
our approximation is to drop the term $(\ddvg - \ddvb)$.
We reserve the expression 'tight coupling' for the validity 
of our assumption that $(\ddvg - \ddvb)$ can be neglected in the slip $\slip$. 
As a criterion, we use $\kt < \frac{2}{10}$ for the photon fluid. When this
threshold is passed, we use Equation \eqref{eqn::dvg} to evolve the photon velocity.
Likewise, for the baryons, we use
${\rm max}(k,\adota)\tc/R < \frac{4}{100}$. Again, when this limit is exceeded, 
we switch to Equation \eqref{eqn::dvb}. In any case, we switch off the approximation
$\Delta \tau = 30 \mpc$ before the first evaluation of the CMB anisotropy sources
(see below).  For a $\Lambda-{\rm CDM}$ model, this is at $\tau \approx 200\mpc$.

To obtain high accuracy \emph{during} tight coupling, it is
crucial to determine $\sg$. Not so much for
the slip \eqref{eqn::slip74}, but more so for the Equations of motion 
\eqref{eqn::tightdv}: the shear reflects
the power that is drained away from the velocity in the 
multipole expansion. This leads to an additional damping for photons.
For the shear, we distinguish two regimes: an early one, where we use a high-order 
analytic approximation and  a later one in which the full multipole equations of motion are used. 

\newcommand{\lo}{l.o.}
\newcommand{\mlo}{\M_2^{l.o.}}
\newcommand{\elo}{E_2^{l.o.}}
\newcommand{\dmlo}{\dot{\M}_2^{l.o.}}
\newcommand{\delo}{\dot{E}_2^{l.o.}}
\newcommand{\sliplo}{\mathcal{\dot V}^{l.o}}
\begin{figure}
\includegraphics[width=0.42\textwidth]{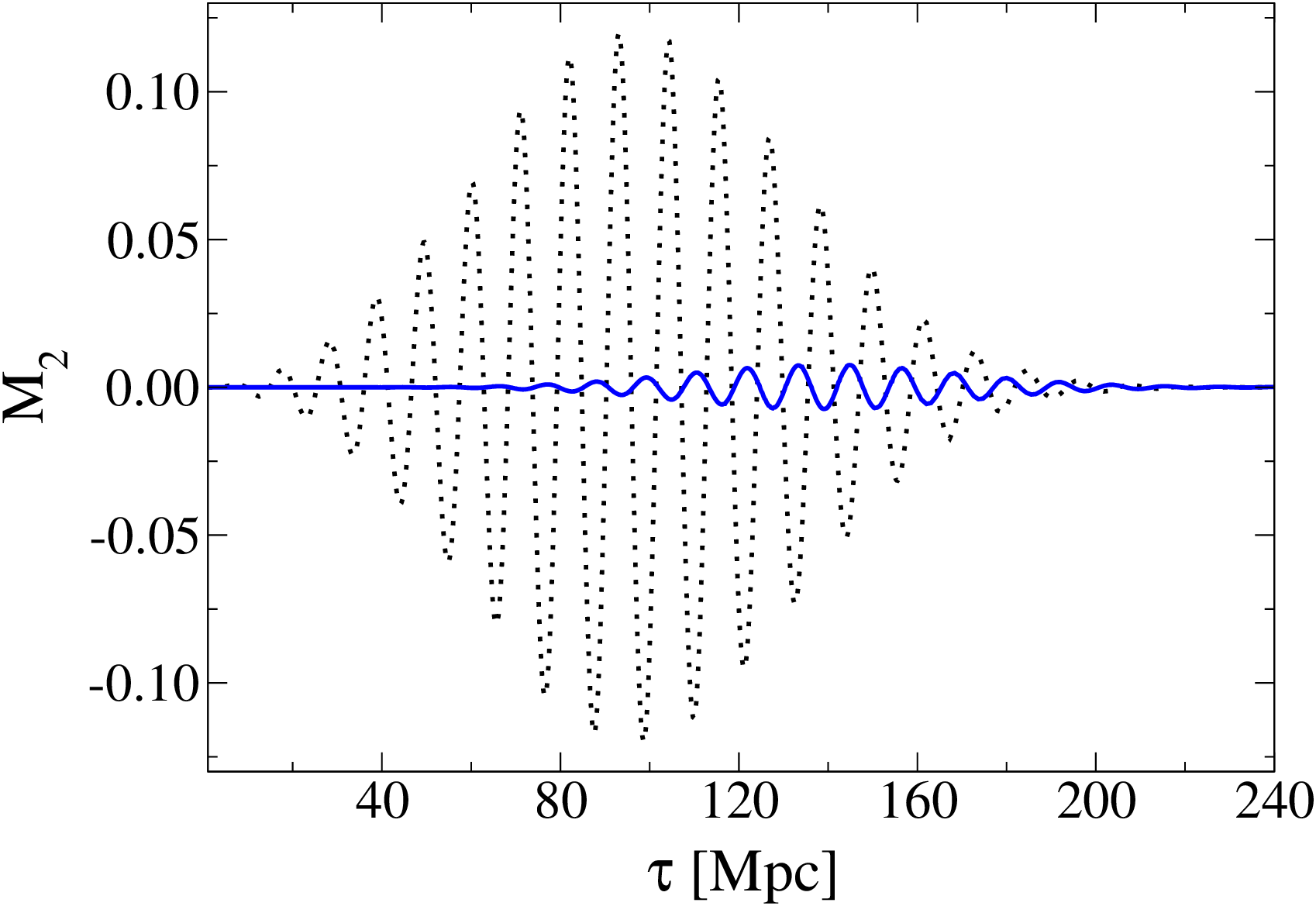} 
\caption{The quadrupole $\M_2$ obtained by a full numerical evolution for a mode of $k=1 \cpm$ (dotted  line).
The solid (blue) line depicts the deviation of our analytic result, Equation \eqref{eqn::m2high} from the 
numerical value. For this mode, we normally switch to the full numerical evolution at $\tau=65 \mpc$ when the 
analytic estimate still holds very well.}\label{fig::quad}
\end{figure}

Since $\tc \ll 1$ at early times, one gets from multiplying
\eqref{eqn::ml}  by $\tc$ that $\M_l \approx \bkt \M_{l-1} l /(2l-1)$.
Hence, higher multipoles are suppressed by powers of $\kt$. 
Approximating this
situation by $\dot \M_3 = \dot E_3 = \M_4 = E_4 = 0$
in Equations \eqref{eqn::ml} and \eqref{eqn::el}, we get
\ea
\M_3 &=&  \frac{3}{5} \bkt \M_2\nonumber \\
E_3 &=& \frac{1}{\sqrt{5}} \bkt E_2. \label{eqn::m3}
\eea
Likewise, we obtain a leading order estimate 
of the quadrupoles by temporarily setting $\dot \M_2 = \dot E_2 = 0$,
\ea 
\frac{5}{2}\sg^{l.o.}=\M_2^{\lo} &=& \frac{8}{9} \bkt \vg\\
E_2^{\lo} &=& -\frac{\sqrt{6}}{4} \M_2^{\lo} \label{eqn::mlo}.
\eea
Inserting Equations \eqref{eqn::m3} 
into the quadrupole Equations \eqref{eqn::m2} and \eqref{eqn::e2}
and using $\dot \M_2 = \dmlo$ and $\dot E_2 = \delo$ as an estimate for the derivative, we get 
the desired expression for the shear
\ee\label{eqn::m2high}
\frac{5}{2}\sg = \M_2 = \frac{8}{9} \kt \vg \left[ 1 - \frac{29}{70} \bkt^2\right] - \frac{11}{6}\tc\dmlo,
\eee
which is precise to order $\tc$ and $\bkt^2$ (see also Figure \ref{fig::quad}). 
The inclusion of the octupole reduces the power of $\M_2$ as expected.

In practice, we use $\mlo$ to calculate the slip $\sliplo$ to leading order. This
in turn is used to calculate $\dvg^{l.o}$. From  $\dvg^{l.o}$, we get
$\dot \mlo$ which in turn is needed to obtain the accurate value of 
$\M_2$ according to Equation \eqref{eqn::m2high}. 
The difference $\Delta\M_2 \equiv \M_2 - \mlo$
is then used to promote $\sliplo \to \slip$ as well as $\dvg^{l.o.} \to \dvg$.
Finally, having $\M_2$ and $\slip$ at hand, we 
get $\dvb$ from  Equation \eqref{eqn::tightdv}.

When this approximation breaks down (sometimes long before
tight coupling ends), we switch to the full 
multipole evolution equations.
Tight coupling is applicable for  $\kt \ll 1$.
Equation \eqref{eqn::m2high} on  one hand goes to higher order in $\kt$,
namely, as $\mlo$ is already of order $\bkt$, our results incorporates quantities
up to $\bkt^3$.
In terms of $\tc$ alone, however, Equation \eqref{eqn::m2high} is  accurate to
order $\tc\bkt$ only. Hence, when $\tc$ reaches $\sim 10^{-1} \mpc$, our analytic expression
is not sufficiently accurate anymore.
This signals the breakdown of our assumption that 
$\dot \M_3 = \dot E_3= 0$ (and likewise for higher multipoles).
Luckily, it is not critical to evolve the full multipole equations even 
when $\tci$ is still substantial.  
This is in strong contrast to the coupled velocity equations which are far more 
difficult to evolve at times when the analytic quadrupole formulae breaks down. 
In essence, distinguishing between tight coupling and the  
treatment of the quadrupole evolution is the key to success here.

\section{A Cure for Rapid Oscillations}

While the gain in speed from the method described in the last section 
is impressive, high $k$-modes would
still require long integration times. To see this, one must consider
the evolution of the photon and neutrino multipole hierarchies.\footnote{
We include the monopole $\delg$ and dipole $\vg$ here.} Our discussion
is aimed at small scale modes which are supposed to be well inside the horizon,
i.e. $k\tau \gg 1$.

Before last scattering, $\bkt \ll 1$ and $\M_l \propto \bkt^{l-1}$ for $l>1$
and so the influence of higher multipoles on $\delg$ and $\vg$
may be neglected to first order. In the small scale limit that we are interested
in, $\delg$ and $\vg$ are oscillating according to $\delg \sim 
\cos(\csg k \tau)$ and $\vb \sim \sin(\csg k \tau)$. As the speed of sound
of the photon-baryon fluid is $\csg \approx \sqrt{1/3}$, we encounter oscillations
with period $\Delta \tau \approx (2 \pi)/(k \csg) \approx 11/k$.
Estimating the time of last scattering with $\tau_{ls} \approx 280 \mpc$, we see
that a mode will perform $\tau_{ls} / \Delta\tau \approx 25 k \mpc$ oscillations
until last scattering. Yet, there are many more oscillations \emph{after} last scattering
which we turn to now.
After last scattering, $\tci$ is negligible and the 
multipole hierarchy of photons effectively  turns into recursion
relations for spherical Bessel functions. The same is true for neutrino multipoles
which roughly evolve like spherical Bessel functions from the start.
Spherical Bessel functions have a leading order behavior similar to 
$j_l(k\tau) \propto (k\tau)^{-3/2}\sin(k\tau)$ for $k \tau \gg 1$ and $k\tau > l$.
The period is then given by $\Delta \tau = (2\pi)/k$. The time 
passed  from last scattering to today, is  $\tau_0 - \tau_{ls} \approx \tau_0 \approx 
14000 \mpc$ for current cosmological models. So we encounter 
$\sim \tau_0 /  \Delta_\tau \approx k \tau_0 /  (2\pi)$
=$2200 \times k \mpc$ oscillations. Numerically, each oscillation necessitates 
$\sim 20 \dots 40$ evaluations of the full set of evolution equations. We therefore
estimate a total of $\sim 6\times 10^{4} \times k \mpc$ evaluations induced
by the oscillatory nature of the solution. So a mode $k = 5\cpm$ needs $\sim 3 \times 10^{5}$
evaluations -- a substantial number.

\begin{figure}
\includegraphics[width=0.4\textwidth]{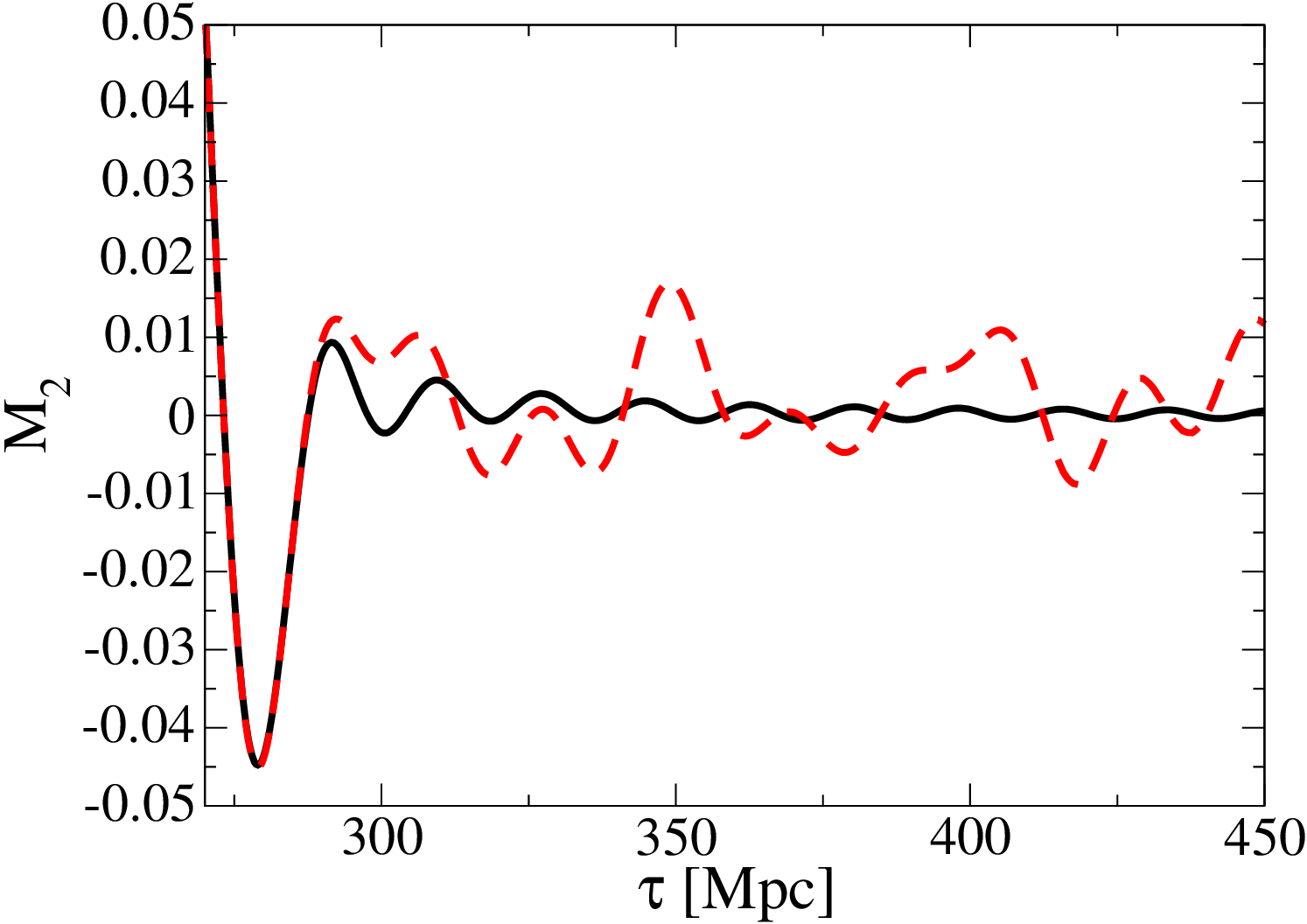} 
\caption{The quadrupole as a function of conformal time $\tau$ for a mode of $k/h=0.5\cpm$ and $h=0.7$. The multipole expansion for photons and
neutrinos has been truncated at $l_{max}=200$ (solid line) and $l_{max}=8$ (dashed line) respectively. In the case
of $l_{max}=8$, power reflected back from the highest multipole $l_{max}$ renders the further evolution of 
the quadrupole unphysical. Indeed the magnitude of the physical oscillations are much smaller than the
reflected ones.
For $l_{max}=200$, reflection effects dominate the evolution from $\tau \sim 1300 \mpc$ on.
In both cases, the effect of shear of realistic particles on the potentials $\phi$ and $\psi$ is negligible by
the time the truncation effects set in.}\label{fig::reflection}
\end{figure}

Since the introduction
of the line-of-sight algorithm, what one really needs for the CMB and
LSS are the low multipoles up to the quadrupoles. In fact, the sources for
temperature and polarization anisotropies are given by
\begin{multline}\label{eqn::source}
S_T = e^{\kappa(\tau)-\kappa(\tau_0)} \left[\dot \phi + \dot \psi \right]
+\dot g\left[\frac{\vb}{k} + \frac{3}{k^2}\dot{\mathcal{C}} \right ] + \ddot g \frac{3}{2k^2} \mathcal{C}\\
+ g\left[\frac{1}{4} \delg + \frac{\dvb}{k} + (\phi+\psi) + \frac{\mathcal{C}}{2} + \frac{3}{2k^2} \ddot{\mathcal{C}}\right]
\end{multline}
and 
\ee\label{eqn::sourceE}
S_E = \frac{3\,g}{2} \mathcal{C}\,\left(k \left[\tau_0 - \tau\right]\right)^{-2}
\eee
Here, $g \equiv \dot\kappa \exp(\kappa(\tau)-\kappa(\tau_0))$ is the visibility with
$\dot\kappa \equiv \tci$ the differential optical depth and $\mathcal{C} \equiv (\M_2 - \sqrt{6} E_2)/10$ 
contains the quadrupole information. The role of  higher multipole moments
is therefore reduced to draining power away from $\delg$, $\vg$ and $\M_2$ and $E_2$ 
(and likewise for neutrinos).
As the oscillations are damped and tend to average out, it suffices to truncate the multipole
hierarchy at low $l \sim 8 \dots 25$ in the line-of-sight approach. This is one of 
the main reasons for its superior speed.
Truncating the hierarchy, though leads to unwanted reflection of power from the
highest multipole $l_{max}$. As one can see in Figure \ref{fig::reflection},
the power reflected back spoils the mono frequency of the oscillations. At
best, the further high frequency evolution of the multipoles is wrong but negligible, because
the oscillations are small and average out. This
is indeed the case in the \CMBFAST/\CAMB/\CMBEASY\ truncation. 
\begin{figure}
\includegraphics[width=0.47\textwidth]{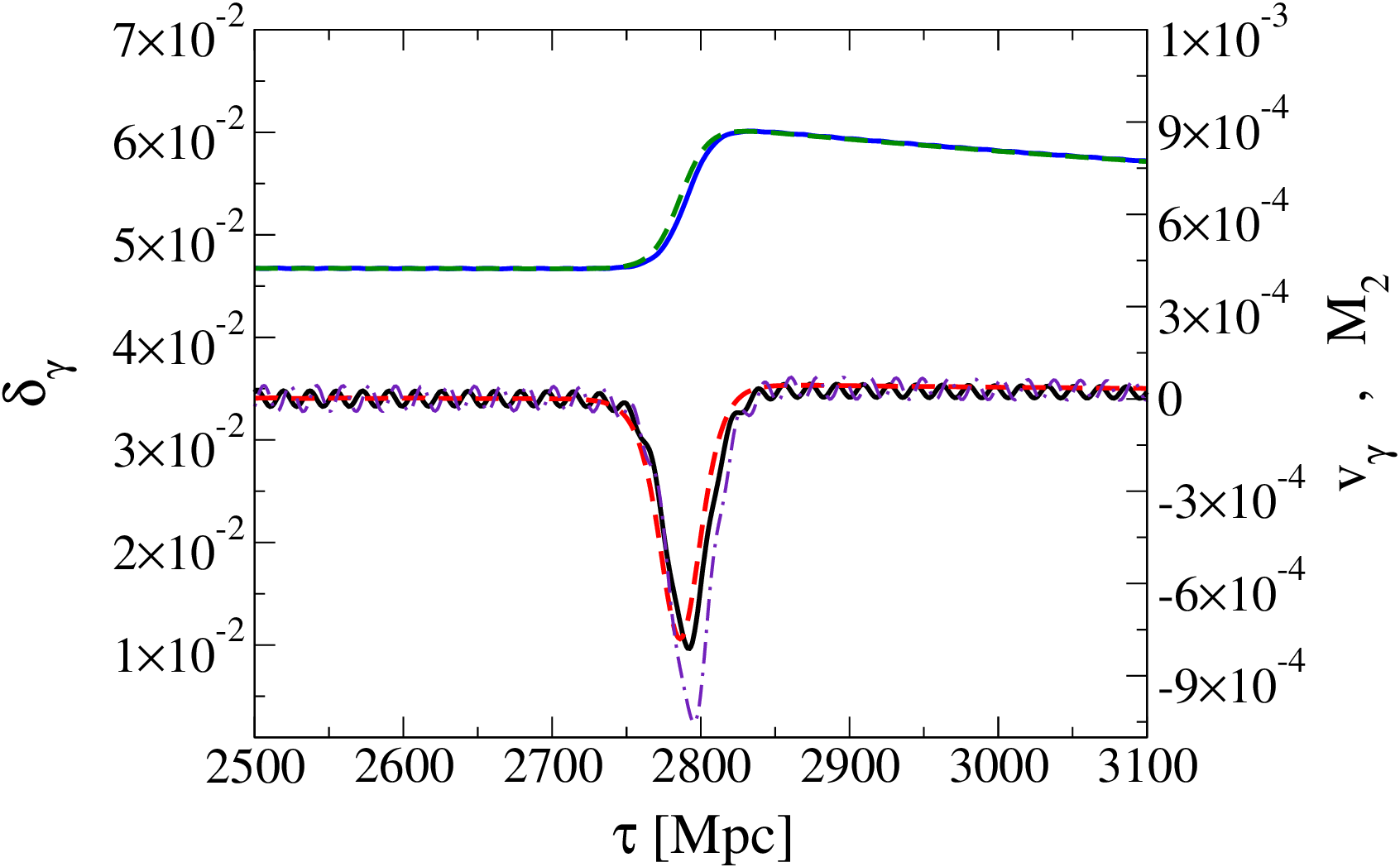}
\caption{Photon density contrast $\delg$ (upper solid [blue] line), $\vg$ (lower solid [black] line)
and quadrupole $\M_2 \equiv (5/2) \sg$ (dashed dotted [indigo] line) as a function of
conformal time $\tau$ before and after re-ionization at $\tau \approx 2800\mpc$.
The [green] upper dashed line is the analytic estimate for $\delg$, Equation
\eqref{eqn::reiondelta} and the lower [red] dashed line is the analytic estimate
for $\vg$, Equation \eqref{eqn::reionv}. The analytic estimate of $\delg$ falls almost
on top of the correct numerical result.
Please note the different scales for $\delg$ and
$\vg$ and $\M_2$ respectively. The quadrupole is roughly of the same order as 
$\vg$. The mode shown is for $k/h=5\cpm$ where $h=0.7$ and  the optical depth to 
the last scattering surface is 
$\tau_{opt} = 0.3$. Please note that we truncated the multipole
hierarchy at sufficiently high $l_{max} = 2500$. With insufficient $l_{max}$,
rapid unphysical oscillations of considerably higher amplitude would be present.}\label{fig::reionization_fine}
\end{figure}

We will now show that the overwhelming contribution from
$\delg$ and $\mathcal{C}$ (and its derivatives) 
of some small scale mode $k > 10^{-1} \cpm$
towards CMB fluctuations  comes from times \emph{before} re-ionization.
To do this, let us find an analytic approximation to
the photon evolution after decoupling and in particular 
during re-ionization. Without re-ionization, and
neglecting $\M_2$ as well as using $\phi \approx \psi$ and 
$\ddot \phi \approx 0$, 
the equation of motions \eqref{eqn::ddelg} and \eqref{eqn::dvg} 
can be cast in the form  
\ee
\ddot{\delta}_\gamma = -\frac{4}{3}k^2 \left(\frac{1}{4} \delta_\gamma + \psi\right),
\eee
which has the particular solution
\ee
\delg = -4 \psi.
\eee
As the oscillations of $\vg$ and higher multipoles are damped roughly
$\propto (k \tau)^{-3/2}$, we see that to good approximation, $\delg = -4\psi$
after decoupling (and before re-ionization) and all higher moments vanish.

During re-ionization, $\tci$ reaches moderate levels again. As $\vb$ has
grown substantial during matter domination, the photon velocity $\vg$ starts
to evolve towards $\vb$. Any increase in magnitude of $\vg$, 
is however  swiftly balanced by a growth of $\delg$ according to Equation 
\eqref{eqn::ddelg}. So roughly speaking, during re-ionization, we may approximate
\ee
0 \approx \dvg \approx \tci \vb + k \left [ \psi + \frac{1}{4}\delg \right ], 
\eee 
where we omit the tiny term $\tci \vg$ and (a bit more worrisome) $\M_2$.
Hence, during re-ionization, the particular solution to
the equation of motion is 
\ee\label{eqn::reiondelta}
\delg \approx -4\psi - 4 \frac{\vb}{k\tc}.
\eee
This approximation holds  well  (see Figure \ref{fig::reionization_fine})
and oscillations on top of it are again damped
and  tend to average out.
Deriving the above \eqref{eqn::reiondelta}, one gets
\ee\label{eqn::reionv}
\vg \approx \frac{3}{k}\left ( 2 \dot\psi - \frac{\dvb}{k\tc} +  \frac{\vb}{k\tc}\frac{\dtc}{\tc} \right).
\eee
Please note that during the onset of re-ionization, $\dtc = 2\adota \tc$
does not hold and it depends on the details of the re-ionization history
to what peak magnitude $\vg$ will reach. Both \CMBFAST\ and \CMBEASY\ implement
a swift switch from neutral to re-ionized and it is  likely that both
serve as upper bounds on any realistic contribution of higher $k$ modes towards
the CMB anisotropies at late time. In other words: as the effects are negligible
for the currently implemented re-ionization history, they will be even more so
for the real one. 
Going back on track, we give an estimate for the amplitude of
 $\M_2$:  assuming $\dot \M_l  \approx 0$ and $\tci \M_l \approx0$,
one gets from the equations of motion \eqref{eqn::ml} that neighboring multipoles 
$\M_l$ are of roughly the same amplitude.  So the amplitude of $\M_2$ and hence
that of the shear $\sg$ is related to that $\vg$, i.e. we find the bound
\ee
\max(|\sg|) \sim \max(|\vg|),
\eee
where it is understood that the maximum is taken of 
full oscillations.
After radiation domination, the metric
potential $\psi$ is given by 
\ee
\psi \sim \frac{a^2 \rho_c \delta_c}{2 \mplank^2 k^2}, 
\eee 
where $\mplank$ is the reduced Plank mass, $\rho_c$ is the energy density of 
cold dark matter and $\delta_c$ is its relative density perturbation.
For modes that enter the horizon during radiation domination, $\delta_c$ is
roughly independent of scale (we omit the overall dependence on 
the initial power spectrum in this argument). Hence, $\psi \propto k^{-2}$ during
matter domination and we see that $\psi \to 0$ and so  $\delg \to 0$ according
to Equation \eqref{eqn::reiondelta}. Provided that $\dtc /\tc$ remains reasonable,
$\vg$ and hence $\M_2$ and $E_2$ will remain negligible as well during re-ionization
and afterwards. 

\begin{figure}
\includegraphics[width=0.41\textwidth]{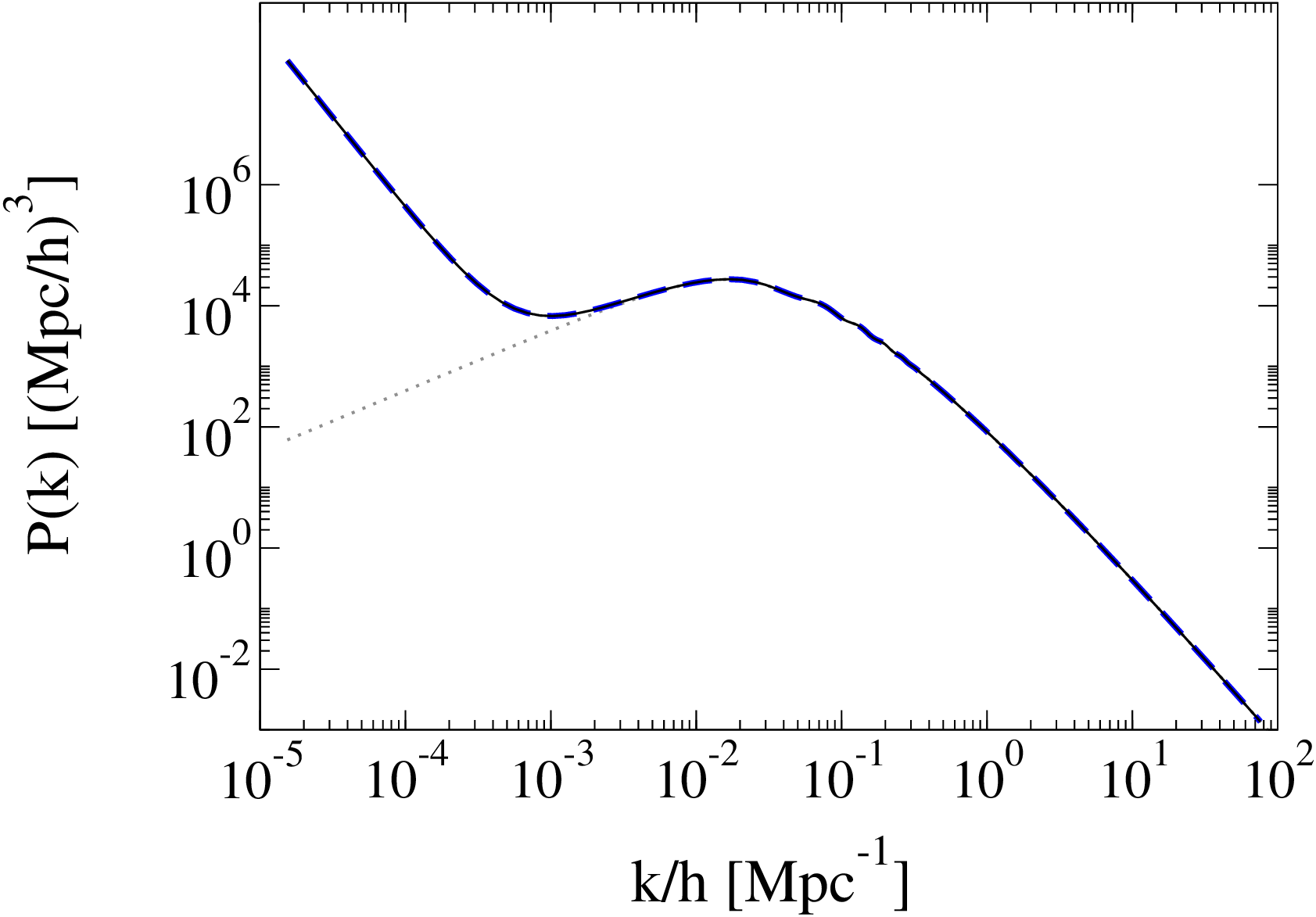}
\caption{Cold dark matter power spectrum using the old gauge invariant implentation (dashed line)
 and  the new strategy in gauge invariant variables (thin solid line). The density contrast
shown is the gauge invariant combination $D_g^{cdm} \equiv \delta^{longit.} - 3 \phi$.
The mean deviation between the curves is $\approx 0.02 \%$. To guide the eye,
we also depict the synchronous gauge power spectrum [thin gray dotted line].
The difference at large scales is due to  gauge ambiguities. Again, we used $h=0.7$.} \label{fig::power}
\end{figure}
For the LSS evolution, neglecting the shear is a good approximation
because Einstein's Equation gives
\ee
\frac{12}{5} a^2 \left [ \bar p_\gamma \M_2 + \bar p_{\nu} \N_2 \right ] = \mplank^2 k^2 (\phi - \psi), 
\eee
where $\N_2$ is the neutrino quadrupole. As $\bar p_{\gamma,\nu} \propto a^{-4}$,
the difference of the metric potentials vanishes for small scale modes, i.e. at least
\ee
(\phi - \psi) \propto (k a)^{-2},
\eee
where we have neglected the decay of the quadrupoles $\M_2$ and $\N_2$ which
give an additional suppression (see also Figure \ref{fig::power}).

As the effect of  $\delg$ and $\M_2$ and $E_2$ at late times for
small scale modes can be neglected (or very well approximated in the
case of $\delg$), we see that there is really no need to propagate
relativistic species at later times.  The key to our final speed up is
therefore to avoid integrating these oscillations after they have
become irrelevant. We do this by multiplying the RHS of 
equations (\ref{eqn::ddelg} - \ref{eqn::el}) as well as the
corresponding multipole evolution equations for relativistic neutrinos
by a damping factor $\Gamma$. Defining $x \equiv k \tau$, we employ
$\Gamma = \{1  - \tanh([x-x_c]/w) \}/2$ with the cross over
$x_c=\max(1000,k\tau_{dilute})$, where $a(\tau_{dilute}) = 5 a_{equ}$
and $a_{equ}$ is the scale factor at matter-radiation equality. This
later criterion ensures that the contribution of relativistic species
to the perturbed energy densities is negligible: from equality on,
$\delta_c \propto a$, whereas $\delg$ decays and $\rho_c/\rho_{rel} \propto a^{-1}$ 
so at least
\ee \delta_c \rho_c : \delg \rho_\gamma \propto a^{-2}, 
\eee 
and similar arguments hold for neutrinos.
Hence, from $\tau_{dilute}$ on, one can safely ignore this
contribution. The former criterion $x_c < 1000$ ensures that
oscillations have damped away sufficiently. The cross-over width $w$ is
rather uncritical. We used $w = 50$ to  make the transition smooth.
Typically, $\tau_{dilute} \approx 400 \mpc$ and one therefore has to follow only a 
fraction of $\tau_{dilute} / \tau_0$ 
oscillations as compared to the standard strategy. 
This corresponds to a gain in efficiency by a factor $\tau_0 / \tau_{dilute} \approx 30$.

To compute the sources $S_T$ and $S_E$, we use the expressions
\ea
\delg &=& \Gamma \delg^{numeric.} - 4 (1-\Gamma)\left[\psi +  \frac{\vb}{k\tc}\right]\\
\C &=& \Gamma \C^{numeric.},  \\
\dot \C &=& \Gamma \dot \C^{numeric.},  \\
\ddot \C &=& \Gamma \ddot \C^{numeric.},
\eea
which interpolate between the numerical value before $\Gamma$-damping and
the analytic approximations, Equation \eqref{eqn::reiondelta} and 
$\C \equiv 0$. Setting $\C \equiv 0$ is an approximation to the small 
value of the quadrupoles averaged over several oscillations. 

For general dark energy models with rest frame speed of 
sound $c_s^2 > 0$ of the dark energy fluid, the dark energy
perturbations well inside the horizon oscillate with high frequency. 
In this case, one needs to suppress the damped oscillations of the 
dark energy fluid perturbations much like those of photons to achieve faster
integration.

\section{Conclusions}
We have improved the integration strategy of modern cosmological Boltzmann
codes. As a first step, we made a conceptual distinction between tight
coupling of the velocities $\vg$ and $\vb$ and the validity of analytic estimates
for the intensity and polarization quadrupole. Doing so allowed us to switch
to the full numerical evolution later. The inclusion of shear
at early times lead to an increase in precision.
In the second part of our work, we  investigated the behavior of photons
after decoupling. We found analytic approximations for both $\delg$ and $\vg$
as well as a  bound on the shear $\sg$.
The contributions of photons and neutrinos towards CMB
anisotropies can be well approximated by using these analytic estimates 
of $\delg$ and $\sg$ for small scale modes deep inside the horizon. 
In fact, for an optical depth $\tau_{opt} \lessapprox 0.2$,
late time effects of photons on the CMB anisotropy sources $S_T$ and $S_E$
may be neglected altogether on small scales. 
We introduced a smooth damping of 
high frequency oscillations of  photon and neutrino multipoles. The damping  effectively
freezes their evolution well inside the horizon.
All in all, our strategy leads to a gain in efficiency of up to factor $\sim 30$ and 
comes close to optimal  accuracy for both the CMB and LSS.

{\bf Acknowledgments} I would like to thank Max Tegmark for drawing my attention to
this subject and  Xue-Lei Chen and Constantinos Skordis for discussions about tight coupling.
This work was supported by 
NSF grant PHY-0099543 at Dartmouth.


\end{document}